\documentclass[twocolumn]{aastex631}

\usepackage{xcolor}
\usepackage{amsmath}
\usepackage{hyperref}
\definecolor{seagreen}{rgb}{0.190, 0.525, 0.361}
\definecolor{salmon}{rgb}{0.98039216, 0.50196078, 0.44705882}

\newcommand{\REV}[1]{{\color{black!100}#1}}

\shorttitle{A CNN to predict Hierarchical Triple Systems Stability}
\shortauthors{Lalande \& Trani}
\graphicspath{{./}{figures/}}

\begin{document}

\title{Predicting the Stability of Hierarchical Triple Systems with Convolutional Neural Networks}

\author[0000-0001-6676-1484]{Florian Lalande}
\affiliation{Okinawa Institute of Science and Technology,
1919-1 Tancha, Onna, Kunigami,
Okinawa 904-0495, Japan}

\author[0000-0001-5371-3432]{Alessandro Alberto Trani}
\affiliation{Research Center for the Early Universe, School of Science,
The University of Tokyo, Tokyo 113-0033, Japan}
\affiliation{Okinawa Institute of Science and Technology,
1919-1 Tancha, Onna, Kunigami,
Okinawa 904-0495, Japan}



\begin{abstract}

\REV{The dynamical stability of hierarchical triple systems is a long standing question in celestial mechanics and dynamical astronomy. 
Assessing the long-term stability of triples is challenging because it requires computationally expensive simulations.}
Here we propose a convolutional neural network model to predict the stability of \REV{equal-mass} hierarchical triples by looking at their evolution during the first $5 \times 10^5$ inner binary orbits. We employ the regularized few-body code \textsc{tsunami} to simulate $5\times 10^6$ hierarchical triples, from which we generate a large training and test dataset. We develop twelve different network configurations that use different combinations of the triples' orbital elements and compare their performances. Our best model uses 6 time-series, namely, the semimajor axes ratio, the inner and outer eccentricities, the mutual inclination and the arguments of pericenter. This model achieves \REV{excellent performance, with an area under the ROC curve score of over $95\%$} and informs of the relevant parameters to study triple systems stability. \REV{All trained models are made publicly available}, allowing to predict the stability of hierarchical triple systems $200$ times faster than pure $N$-body methods.

\end{abstract}




\section{Introduction}
\label{sec:introduction}


The gravitational three-body problem is one of the oldest issues in astronomy and classical mechanics. Despite being studied since the time of Newton himself, no closed-form solution of the three-body problem exists to date, and we are limited to understanding it using numerical or statistical approaches \citep{,monaghan1976a,monaghan1976b,nash1978,mikkola1994,valtonen2006,trani2019b,stone2019,manwadkar2020,manwadkar2021,kol2021}.


Besides its importance in mathematical physics, the three-body problem has numerous applications in modern astrophysics. For example, the hierarchical three-body problem, in which a binary is orbited by a distant third companion, plays an important role in shaping the architectures of planetary systems \citep[e.g.][]{munoz2016}, and in the formation of gravitational wave sources observable by LIGO-Virgo-KAGRA \citep[e.g.][]{hoang2018,trani2022}.

Modeling hierarchical triples is computationally challenging, even when considering purely Newtonian gravitational forces. The most straightforward approach is direct numerical integration of the Newtonian equations of motion. However, long-term integration is computationally expensive, because the integration timestep needs to be short enough to resolve the motion of the inner binary. In addition, the inevitable accumulation of numerical errors makes individual simulations unreliable, and requires running several realizations of the same system in order to faithfully predict its evolution \citep{portegieszwart2014}. Fast approximate techniques like the double-averaging approximation can be used in alternative to direct $N$-body integration, however these have limited validity regimes \citep{marchal1990,krymolowski1999,ford2000,breiter2005,tremaine2009,naoz2013a,luo2016,grishin2017}. In particular, this approach is based on the assumption of hierarchical stability, i.e. that the system will remain in a hierarchical state forever and no energy exchange between the inner binary and the tertiary object occurs.

In fact, the hierarchical state of some triples might only be temporary, with dynamical instability manifesting even after thousands of orbital periods of apparent stability. Once a triple becomes unstable, the hierarchy is disrupted and the three bodies undergo violent and chaotic interactions, until one body is ejected and only a binary is left. The mechanisms by which some triples become unstable are still not fully understood \citep{toonen2021,hayashi2022}, yet many authors have tried to find analytic criteria to assess the stability of hierarchical triples \citep[][]{mardling2001,Mushkin2020,hayashi2022}.

On a similar problem, \citet[][]{tamayo2020} have developed SPOCK, the Stability for Planetary Orbital Configurations Klassifier. This is a machine-learning model capable of predicting the stability of compact multi-planet systems. SPOCK uses a gradient boosting classifier on 10 specific features computed onto the first $10^4$ inner orbits, and tries to predict the system stability after $10^9$ orbits.

In this work, we train an artificial neural network using the time-series of the Keplerian elements in order to predict the stability of hierarchical triple systems. Convolutional neural networks \citep[CNNs, ][]{lecun1989} have notably shown great performances for time-series classifications \citep{Fawaz2019}, and we therefore develop a CNN tailored for hierarchical triple systems stability prediction. We make the trained models and the source code available as an open-source package.

Section~\ref{sec:methods} introduces the methods employed in this work, including the dataset generation, the training data selection and the design of the CNNs. Section~\ref{sec:results} presents the classification results of the networks and the diagnosis of the selected best model. Finally, Section~\ref{sec:discussion} addresses possible limitations and summarizes our work.


\section{Methods}
\label{sec:methods}

In this section we first describe the generation of the dataset of hierarchical triples (Section~\ref{subsec:dataset_gen}); next, we introduce the set of Keplerian elements and their transformations used as input parameters for training (Section~\ref{subsec:input_data}); and we finally provide the description of the various CNN architectures as well as the training strategy (Section~\ref{subsec:cnns}).

\subsection{Dataset of hierarchical triple systems}
\label{subsec:dataset_gen}

We use \textsc{tsunami} \citep{Trani2019}, a modern, regularized code suited to evolve few-body self-gravitating systems. The algorithm can handle close encounters with extreme accuracy thanks to a combination of numerical techniques, namely: algorithmic regularization \citep{mikkola1999}, chain coordinate system \citep{mikkola1993} and Bulirsch-Stoer extrapolation \citep{bulirschstoer2002}.

In the algorithmic regularization scheme, we solve the equations of motion using a second order leapfrog derived from a transformed Hamiltonian extended in phase space. In the resulting kick-drift-kick scheme, time becomes an integrated variable, just like positions and velocities, because the system is advanced along a new independent variable, that can be considered as a fictional time. The advantage of this algorithm is that the integration timestep can remain large even for collision orbits, where the acceleration going to infinity would cause numerical problems in traditional integration schemes. Because the leapfrog scheme is only second-order, we increase the accuracy of the integration by employing Bulirsch-Stoer extrapolation with an adaptive timestep. 
The chain coordinate system is instead a coordinate system alternative to the more common center-of-mass coordinates, which serves to reduce round-off errors when two close particles are very far from the center of mass. 
\textsc{tsunami} comes with a Python interface, which makes it particularly convenient for the generation of our dataset of hierarchical triple systems.

The hierarchical triple systems we consider comprise three bodies of $10 \rm\,M_\odot$, which we refer to as $1$, $2$ and $3$. Bodies $1$ and $2$ form the inner binary of the hierarchical system. The outer orbit is composed of the inner binary center of mass and body $3$. The initial parameters of the system are sampled such that we observe both long-lived systems and disruptions.

We sample the initial inner eccentricity with the square of a uniform distribution
($e_1 \sim \mathcal{U}[0,1]^2$)
and uniformly sample the initial outer eccentricity 
($e_2 \sim \mathcal{U}[0,1]$). The distribution for the inner binary eccentricity was inspired by the $e_1$ distribution of triples formed in stellar cluster simulations \citep{trani2022}.
\REV{We sample the initial inner semi-major axis from a log-uniform disribution between 1 and 100 au
($\log a_1/{\rm au} \sim \mathcal{U}[\log 1, \log 100]$)}.
We generate the initial mutual inclination by uniformly sampling the arccosine of the angle 
($\cos{i_{\rm mut}} \sim \mathcal{U}[-1, 1]$).
Then, we also uniformly and independently sample both arguments of periapsis ($\omega_1, \omega_2 \sim \mathcal{U}[-\pi, \pi]$). Next, we sample the initial mean anomaly for the inner and outer binaries uniformly ($M_1, M_2 \sim \mathcal{U}[0, 2\pi]$), and calculate the true anomalies ($\nu_1, \nu_2$) accordingly.

Finally, we make use of the numerical stability criterion of \citet{mardling2001} to compute a theoretical limit for the semi-major axis of the outer binary $a_2^{\rm{lim}}$. \citet{mardling2001} proposed a simple expression to assess the stability of hierarchical triple systems, empirically derived from simulations. This criterion aims at providing a boundary between stable and unstable initial configurations for hierarchical triple systems. According to their criterion, the triple is stable if the outer semi-major axis $a_2$ is greater than
\begin{multline}
a_2^{\rm{lim}} = 2.8 \left( \frac{(m_1 + m_2 + m_3) (1+e_2)}{(m1 + m_2) (1-e_2)^3} \right)^{2/5} \\ \times \left( 1 - \frac{0.3 i_{mut}}{\pi} \right) a_1 
\end{multline}

\begin{figure}[ht]
    \centering
    \includegraphics[width=0.45\textwidth]{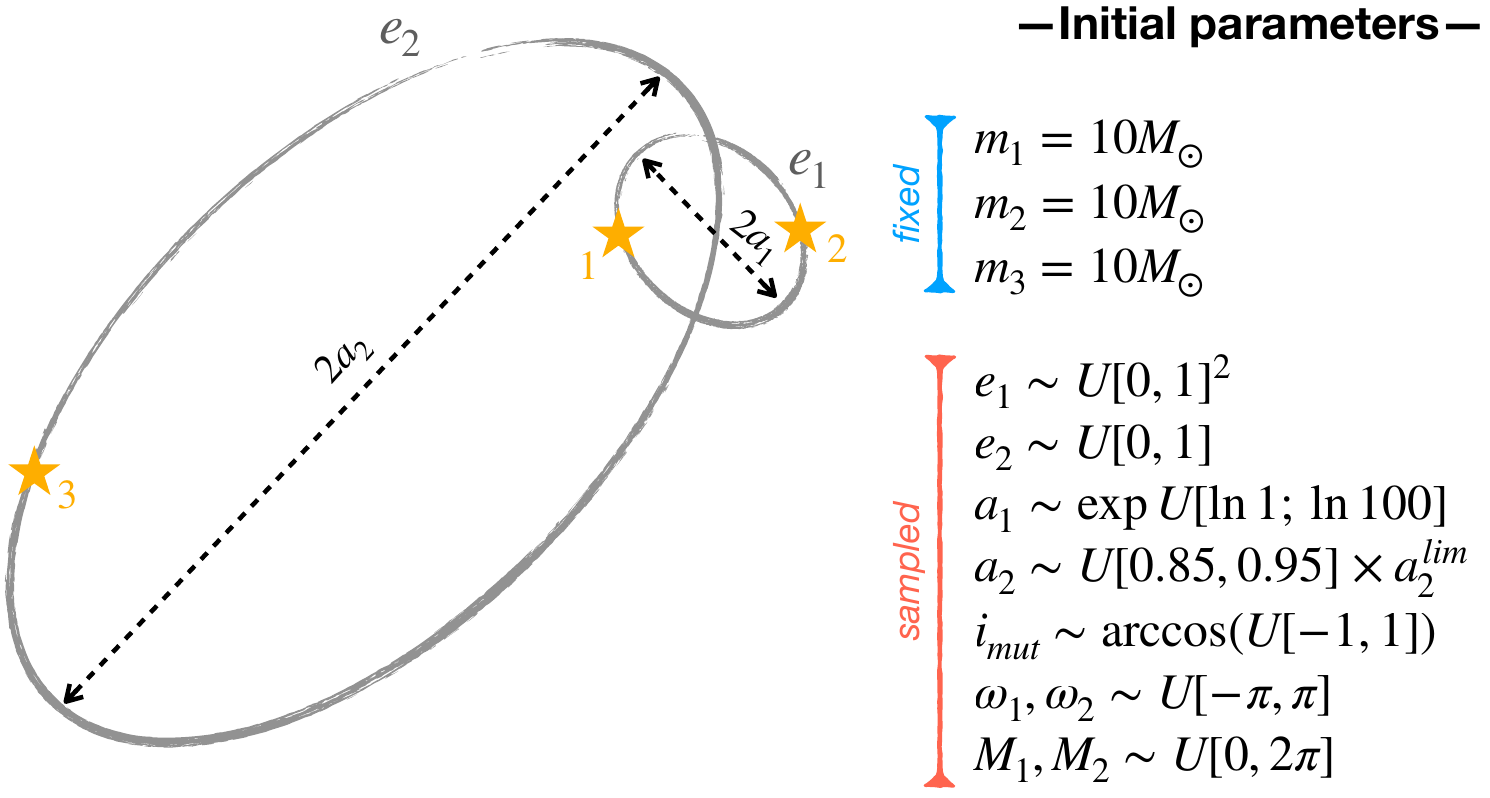}
    \caption{Initial configuration and input parameters for the hierarchical triple systems simulated with \textsc{tsunami}. The masses are fixed (in blue) for all simulations while all other parameters are sampled (in red) to explore the initial parameter space. The outer semi-major axis $a_2$ is sampled such that we expect medium-lived and long-lived systems, while avoiding obviously stable configurations.}
    \label{fig:initial_setup}
\end{figure}

For our simulations, we uniformly sample the initial semi-major axis of the outer binary $a_2$ in a range slightly below $a_2^{\rm{lim}}$: $a_2 \sim \mathcal{U}[0.85, 0.95] a_2^{\rm{lim}}$. \REV{This sampling strategy prevents from simulating obviously stable or unstable triple systems.} The uniform distribution range of $[0.85, 0.95]$ has been chosen empirically, such that we obtain both long-lived and medium-lived systems in a reasonable amount of computational time. In the most unstable scenario, we have $a_2 = 0.85 a_2^{\rm{lim}}$ and the system is likely to undergo fast disruption; while in the most stable scenario, we have $a_2 = 0.95 a_2^{\rm{lim}}$ and the simulation is expected to last longer. Very short-lived simulations are not computationally expensive and can simply be discarded. \autoref{fig:initial_setup} summarizes the initial configuration and input parameters sampling.

We run each simulation for a maximum time of $t_{\rm f} = 10^8 P_1$, where $P_1$ is the initial inner orbital period. Simulations are stopped when either $t_{\rm f}$ is reached (stable simulation) or a breakup happens (unstable simulation), whatever time is the shortest. A breakup is defined as the moment when the semimajor axes ratio $a_1/a_2$ deviates from the initial value by more than 15\%.
\REV{Strictly speaking, breakup usually refers to when a triple breaks into an binary and a single body, escaping on an unbound hyperbolic trajectory. However, before such a breakup happens, the system enters a chaotic evolution phase, which in the context of the non-hierarchical three-body problem is called ``resonant interaction'' \citep[e.g.][]{hut1983a}. During this phase, the hierarchy of the system is already broken and dynamically unstable, but the system might still take a long time to finally decay into an unbound-binary single \citep[e.g.][]{manwadkar2020}. For this reason, we adopt the above condition on the semimajor axes ratio as our definition of instability. While some oscillations of the semimajor axis ratio can happen over small timescales, we noticed that after $a_1/a_2$ changes significantly (about 15\%), the energy transfer between inner and outer orbits cannot stop anymore, and the triple will eventually break.

It is worth noting that our criterion for instability ignores the effects of the secular evolution of triples, such as the von~Zeipel-Kozai-Lidov mechanism \citep{zeipel1910,koz62,lid62}, which causes long-term oscillations in mutual inclination and inner binary eccentricity of highly-inclined triples. We do not regard the secular evolution as an instability itself, but as a process that can trigger the instability in terms of energy exchange between inner and outer orbit. For example, the inner binary eccentricity may increase due to the von Zeipel-Kozai-Lidov mechanism, which then causes the inner binary to become more perturbed \citep[see sec.~6 of][]{hayashi2022}. Eventually, this will manifest in a change of the semi-major axis ratio. The reciprocal is not always verified: secular evolution does not necessarily trigger the dynamical instability, and the triple may proceed evolving in a stable configuration.
}

During simulations, we periodically sample the following 8 parameters: current time in years $t$, inner orbit eccentricity $e_1$, outer orbit eccentricity $e_2$, inner orbit major-axis $a_1$, outer orbit major-axis $a_2$, mutual inclination between inner and outer orbits $i_{\rm mut}$, inner orbit argument of periapsis $\omega_1$ and outer orbit argument of periapsis $\omega_2$. These parameters are recorded every $\Delta t=5 \pi P_1$, from the beginning of the simulation until $5 \times 10^5 P_1$, that is 0.5\% of the total simulation duration. 
\REV{This arbitrary cutoff at 0.5\% is a tradeoff between accuracy, which increases by proving longer time-series, and computational efficiency, which naturally decreases for higher cutoffs.}
The relevant data corresponds to seven time series of following orbital elements $(e_1, e_2, a_1, a_2, i_{\rm mut}, \omega_1, \omega_2)$.
\REV{Note that these seven attributes completely describe the secular evolution of the system. Therefore, other quantities like angular momentum or orbital energies can be derived from these seven orbital elements or their combination. If any of these derived quantities is fundamentally important for our classification problem, we expect the network to find them automatically.}

We simulate the evolution of 5,000,000 triple hierarchical systems with \textsc{tsunami}. Since we are interested in the analysis of the first 0.5\% of the maximum simulation lifetime, we discard all simulations that underwent disruption before $5 \times 10^5 P_1$. This corresponds to about 90\% of our simulations. 
\REV{Note that this process may introduce a bias in our dataset because the neural network will not be trained on very short-lived (${<} 5 \times 10^5 P_1$) simulations. Including all 5,000,000 simulations would have led to artificially high classification performances as the very short-lived simulations break up too quickly. Since very short-lived systems can be simulated until disruption at a negligible computational cost, we decided to focus on the critical simulations that develop the instability much later in time, and are more complicated to classify. Training an artificial neural network on imbalanced datasets remains a complex open problem \citep{Fernandez2018}.}

\begin{figure}[ht]
    \centering
    \includegraphics[width=0.45\textwidth]{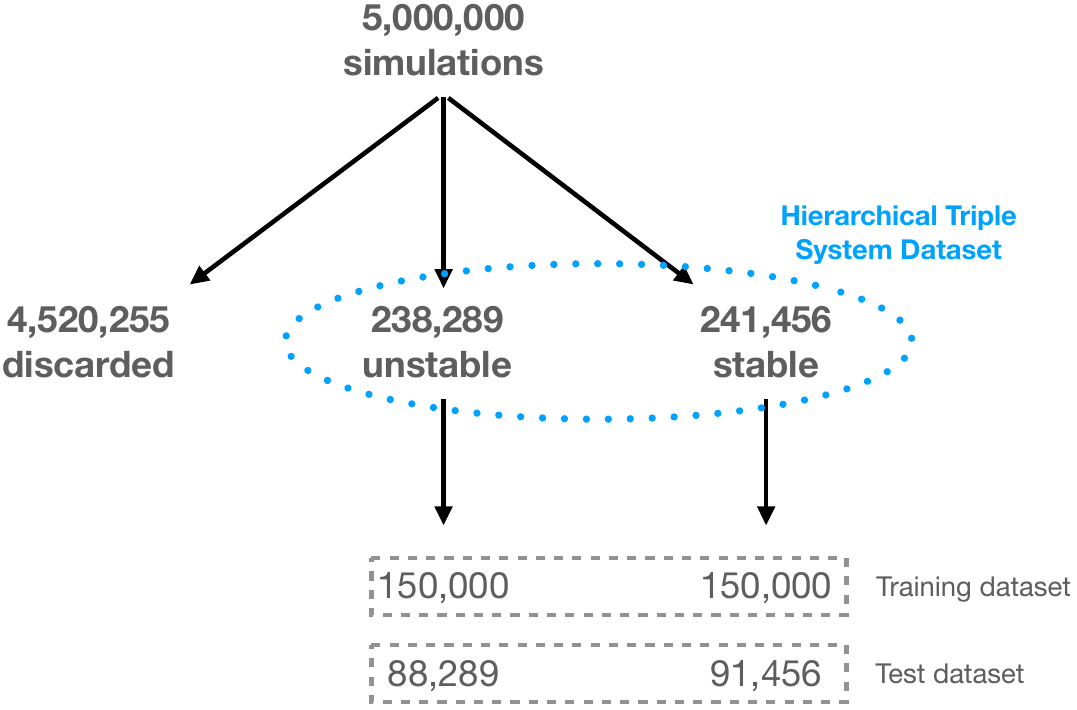}
    \caption{Simulated dataset and training/test split. The simulations that did not last longer than $5 \times 10^5 P_1$ ($0.5\%$ of our final time) have been discarded. The remaining simulations have been divided into two classes: stable and unstable. Stable simulations did reach the maximum final time $t_{\rm f} = 10^8 P_1$ without disruption, while unstable simulations did not. In each class, we selected $150,000$ simulations for training, the rest being kept for testing.}
    \label{fig:dataset}
\end{figure}

About 5\% of our simulations were found medium-lived and underwent disruption between $5 \times 10^5 P_1$ and $10^8 P_1$. These 238,289 simulations are included in the dataset and labeled as ``unstable''. Finally, 241,456 simulations (another 5\% of the total) reached the final time $t_{\rm f}$ without disruption. We include these simulations in the dataset as ``stable''.
\REV{We purposefully discarded the simulations for which $t_{\rm f} < 100 \, P_2$, because the third body does not have enough time to interact with the inner binary, making the simulations appear artificially stable.}
As a result, our hierarchical triple systems dataset is comprised of 479,745 simulations, with approximately half being stable and the other half unstable. \autoref{fig:dataset} shows a summary of the simulation results and the hierarchical triple systems dataset generation.

\subsubsection{Comparison with Mushkin \& Katz estimate}

\begin{figure}[ht]
    \centering
    \includegraphics[width=0.45\textwidth]{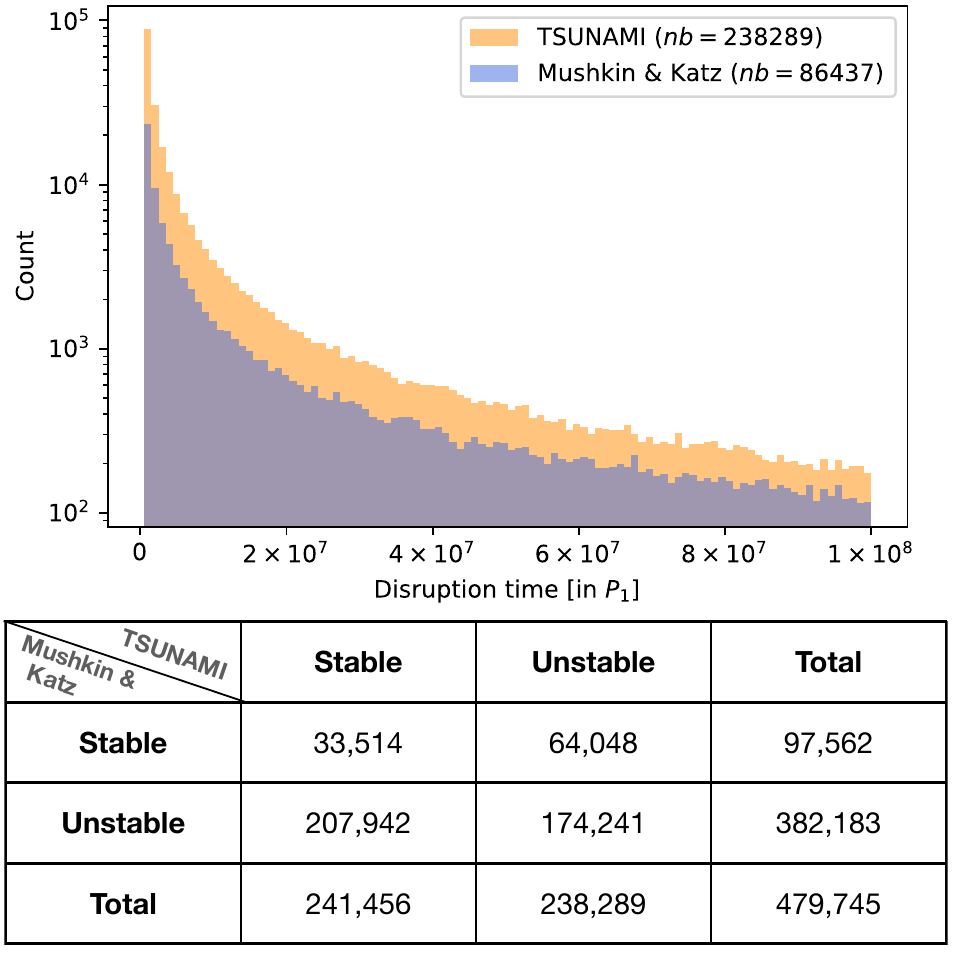}
    \caption{Disruption time distribution for \textsc{tsunami} unstable simulations and comparison with the \citet[][]{Mushkin2020} disruption time estimates. The y-axis is logarithmic: our unstable dataset (in orange) is very biased towards short-lived simulations. The blue histogram displays the distribution of disruption time estimates following \citet[][]{Mushkin2020}, where the predictions below $5 \times 10^5 P_1$ have been removed ($382,183 - 295,746 = 86,437$). The contingency table shows the comparison over all simulations: the approximation of \citet[][]{Mushkin2020} overestimates instability.}
    \label{fig:disruption_hist_table}
\end{figure}

It is worth noting that the unstable simulations of our dataset are strongly biased towards short-lived systems. As can be seen in \autoref{fig:disruption_hist_table}, the disruption time distribution of our \textsc{tsunami} unstable simulations decreases exponentially.

For comparison, we computed the estimated disruption time of hierarchical triple systems following \citet{Mushkin2020}. Their estimate was derived after modeling the disruption of hierarchical 3-body systems by a random-walk process. \citet{Mushkin2020} estimate the number of inner orbital periods $N$ before a triple disrupts as
\begin{multline}\label{eq:mushkin}
N = (1 - e_2)^2 \left( \frac{a_2(1-e_2)}{a_1} \right)^{-7/2} \left( \frac{(m_1 + m_2)^2}{m_1 m_2} \right)^2 \\
\times \left( \frac{m_1 + m_2 + m_3}{m_1 + m_2} \right)^{5/2} \\
\times \exp{ \left( \frac{4 \sqrt{2}}{3} \sqrt{\frac{m_1 + m_2}{m_1 + m_2 + m_3}} \left( \frac{a_2(1-e_2)}{a_1} \right)^{3/2} \right) }
\end{multline}
which for our equal mass systems with $m_1 = m_2 = m_3 = 10 \,\rm M_\odot$ reduces to
\begin{multline}
N = \frac{36 \sqrt{3}}{\sqrt{2}} \left( \frac{a_2}{a_1} \right)^{-7/2} (1-e_2)^{-3/2} \\
\times \exp{ \left(\frac{8}{3\sqrt{3}} \left( \frac{a_2}{a_1} \right)^{3/2} (1-e_2)^{3/2} \right)}
\end{multline}

The estimated disruption time of our $479,745$ simulations are also following an exponential decay (see the blue histogram of \autoref{fig:disruption_hist_table}). The contingency table at the bottom of \autoref{fig:disruption_hist_table} compares the true stability (obtained with \textsc{tsunami}, in orange) with the estimated stability (i.e. disruption time greater than $10^8 P_1$ following \citet{Mushkin2020}, in blue). Amongst the disruption time estimates below $10^8 P_1$, most of them are below $5 \times 10^5 P_1$ and are therefore not displayed in \autoref{fig:disruption_hist_table}.

Interestingly, the majority of \textsc{tsunami} unstable simulations ($73.1\%$) have a predicted disruption time below $10^8 P_1$ and are therefore correctly classified as unstable simulations by \citet[][]{Mushkin2020}. However, only $13.9\%$ of the stable \textsc{tsunami} simulations have a predicted disruption time above $10^8 P_1$ using the same criterion. These results suggest that the predictions of \citet[][]{Mushkin2020} tend to overestimate instability. 

It is worth noting that this comparison is only valid in a statistical sense. Understandably, the \citet[][]{Mushkin2020} criterion fails to predict the disruption time of individual simulations. This occurs because \autoref{eq:mushkin} is a simplified expression that does not take into account other variables like the true anomalies or the apsidal orientations, which can critically affect the long-term evolution of triples \citep{hayashi2022}.

\subsection{Input channels, training and test data}
\label{subsec:input_data}

The time series of the Keplerian orbital elements are not presented as they are to the artificial neural network. Instead, we perform intermediate pre-processing to standardize the data and facilitate the task of the neural network. \autoref{tab:input_channels} introduces the formatting of the time series that has been performed.

\begin{table}[ht]
\centering
\begin{tabular}{|c|c|c|}
\hline
\textbf{Channel number} & \textbf{Value}                   & \textbf{Range}     \\ \hline
\texttt{0}              & $e_1$                            & $[0, 1]$           \\ \hline
\texttt{1}              & $e_2$                            & $[0, 1]$           \\ \hline
\texttt{2}              & $a_1$                            & $\mathbb{R}^+$     \\ \hline
\texttt{3}              & $a_2$                            & $\mathbb{R}^+$     \\ \hline
\texttt{4}              & $\frac{i_{\rm mut}}{\pi}$            & $[0, 1]$           \\ \hline
\texttt{5}              & $\frac{\omega_1 + \pi}{2 \pi}$   & $[0, 1]$           \\ \hline
\texttt{6}              & $\frac{\omega_2 + \pi}{2 \pi}$   & $[0, 1]$           \\ \hline
\texttt{7}              & $\frac{a_1}{a_2}$                & $\mathbb{R}^+$     \\ \hline
\texttt{8}              & $\frac{a_1(1+e_1)}{a_2(1-e_2)}$  & $\mathbb{R}^+$     \\ \hline
\end{tabular}
\caption{Description of the time series pre-processing leading to the network input channels. The raw time series are standardized to facilitate the information processing of the CNNs. Various combinations of the presented input channels are fed into the CNNs for classification.}
\label{tab:input_channels}
\end{table}

\begin{figure}[ht]
    \centering
    \includegraphics[width=0.47\textwidth]{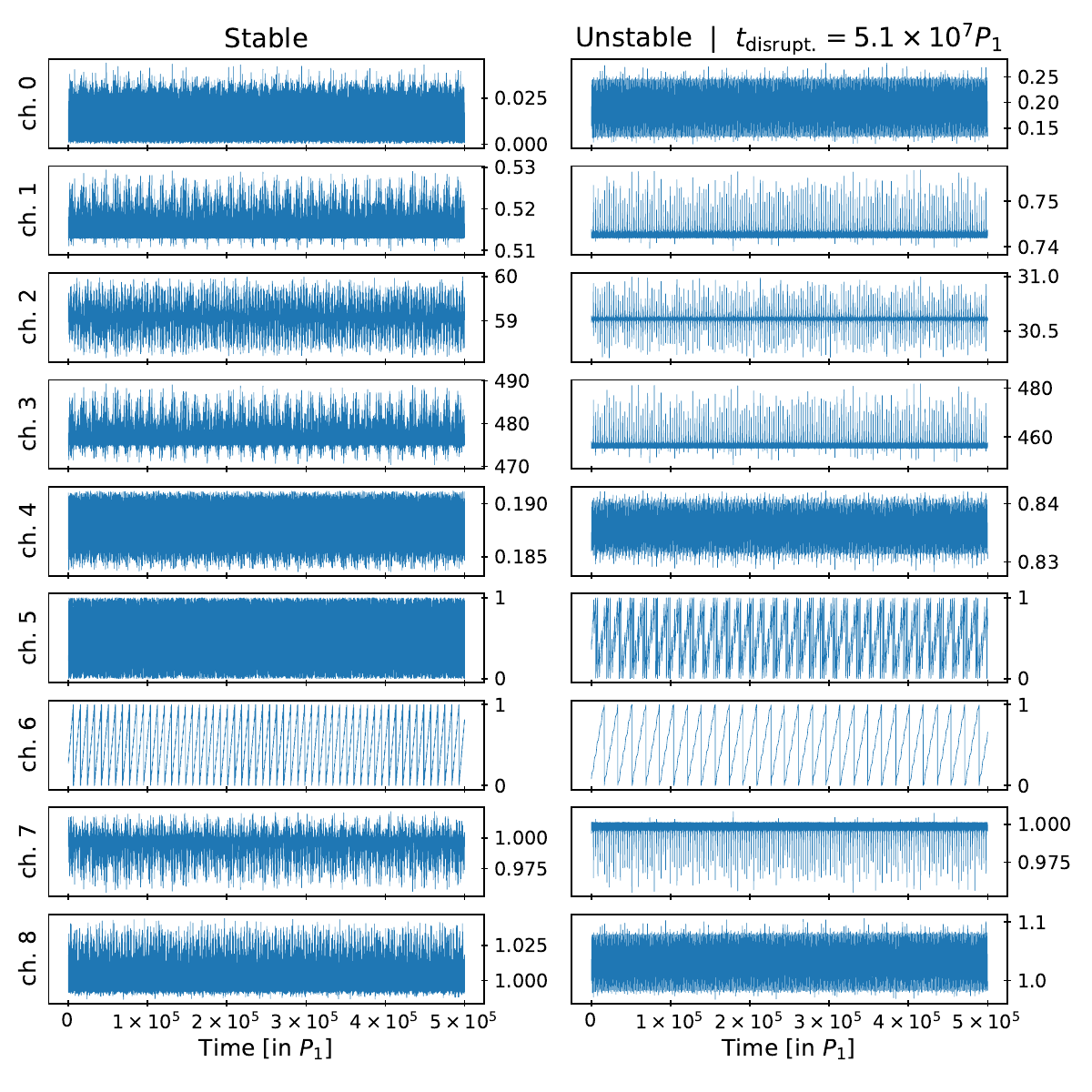}
    \caption{Instance of all available input channels. The left panel corresponds to a stable simulation in our dataset, while the right panel shows an unstable simulation which undergoes disruption after $5.1 \times 10^7 P_1$. Note the scale on the y-axis for channels 2 and 3 (semi-major axes): there magnitudes greatly vary in the dataset as they are not normalized. Channels 2 and 3 are not presented to the network, we use channels 7 and 8 instead.}
    \label{fig:data_example}
\end{figure}

First, the mutual inclination $i_{\rm mut}$ and the arguments at periapsis $\omega_1$ and $\omega_2$ are scaled by $\pi$ and $2 \pi$ respectively to lie in the interval $[0, 1]$. Second, we compute derived quantities from the semi-major axes $a_1$ and $a_2$. 
The ratio of semi-major axes $a_1 / a_2$ and the ratio 
\begin{equation}\label{eq:peridist}
	\frac{a_1(1+e_1)}{a_2(1-e_2)}.
\end{equation}
This second quantity, which we term apsidal separation ratio, is the ratio of the inner binary apocenter distance to the outer binary pericenter distance, and estimates how well the two orbits are separated. Both quantities are dimensionless and normalized to 1 at $t=0$. \autoref{fig:data_example} displays all available input channels for a random simulation in each class.

Because some channels are redundant, we select only a subset of input channels for the CNN. Six sets of input channels are tested:
\begin{itemize}
    \item Channels \{0, 1, 4, 7\} as benchmark.
    \item Channels \{0, 1, 4, 5, 6, 7\} to assess the importance of the arguments of periapsis.
    \item Channels \{0, 1, 4, 8\} to compare the ratio of semi-major axes to the apsidal separation ratio.
    \item Channels \{0, 1, 8\} to assess the importance of the mutual inclination.
    \item Channels \{4, 7\} to evaluate the network performances in the absence of the eccentricities.
    \item Channels \{4, 8\} to check whether the information on the eccentricities encoded in the apsidal separation ratio can be enough.
\end{itemize}

The hierarchical triple system dataset is comprised of $241,456$ stable simulations and $238,289$ unstable simulations. We create a training dataset of $150,000$ stable and $150,000$ unstable simulations. We use all remaining simulations -- that is $91,456$ stable and $88,289$ unstable simulations -- for testing. The lower part of Figure \ref{fig:dataset} shows the split of the original dataset between training and test datasets. Having slightly more stable than unstable simulations in the test dataset is not an issue when testing.

\subsection{CNN designs}
\label{subsec:cnns}

Convolutional Neural Networks (CNNs) are a type of artificial neural networks tailored for image processing. They are translation invariant thanks to the use of convolutions with trainable filters sliding along the input data \citep[][]{lecun1989}.

In spite of their traditional use for 2-dimensional image data, CNNs can well be applied for 1-dimensional time series data processing. CNNs have been shown to provide with state-of-the-art performances for time series classification \citep[][]{Fawaz2019}. For this reason, we decide to develop two CNN models for binary classification of time series.

We introduce two CNN model architectures to account for different dependencies between input channels. Architecture A and Architecture B are presented in \autoref{app:cnn_architectures}. The activation functions between the network layers are presented in \autoref{app:activation_functions}.
\REV{Both network architectures were inspired from well-established CNNs for classification tasks \citep[like Alexnet or VGG16][]{Russakovsky2015,Krizhevsky2017}. We adapted their 2-dimensional architectures to match our time-series input data.}

All CNNs are trained with the same strategy, regardless of the set of input channels used. We use the \texttt{Keras} library \citep[][]{Chollet2015} for \texttt{TensorFlow} \citep[][]{Abadi2016} in \texttt{Python}. We use the Adam optimizer \citep[][]{Kingma2015} to minimize the binary cross-entropy with respect to the trainable parameters during training. Details for the training strategy are provided in \autoref{app:training_strategy}.

\begin{figure*}[ht]
    \centering
    \includegraphics[width=1\textwidth]{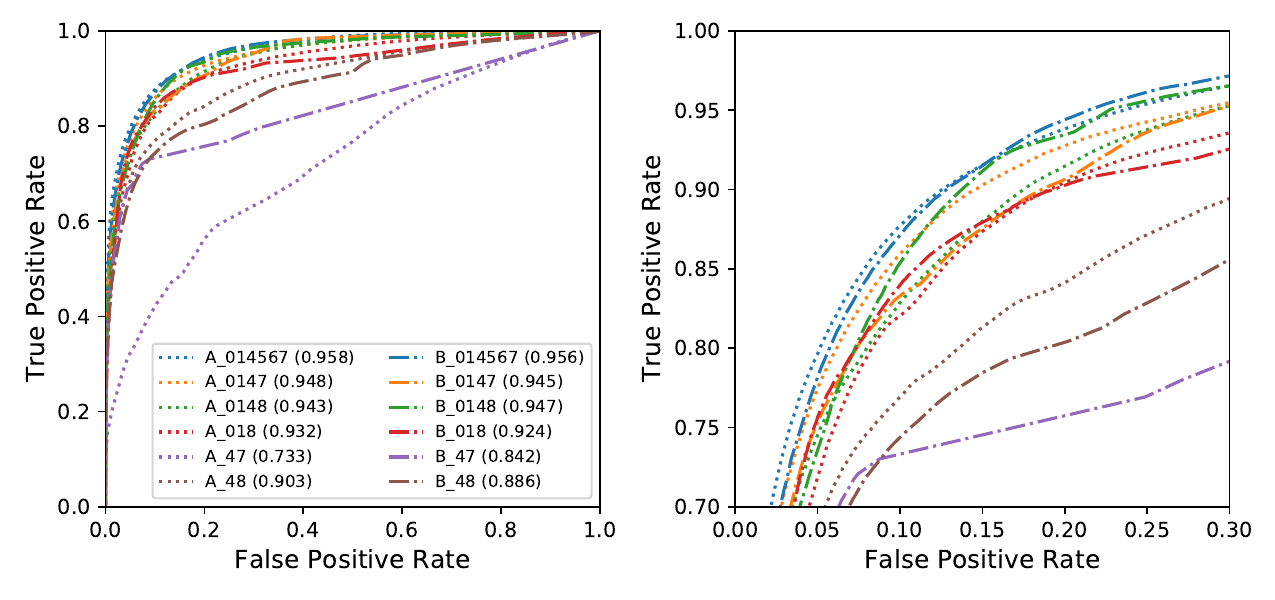}
    \caption{ROC curves of the twelve configurations under study. Not providing the eccentricities $e_1$ and $e_2$ to the CNN (channels \texttt{0} and \texttt{1}) leads to much poorer performances. The \textbf{(right)} panel is a closer look at the upper-left corner of the \textbf{(left)} panel. CNN Architecture B performs as good as Architecture A, in spite of less trainable parameters. The mutual eccentricity $i_{\rm mut}$ (channel \texttt{4}) and the arguments at periapsis $\omega_1$ and $\omega_2$ (channels \texttt{5} and \texttt{6}) appear to bear additional meaningful information.}
    \label{fig:roc_curve}
\end{figure*}


\section{CNN predictions}
\label{sec:results}

As we use two CNN architectures with six sets of input channels for the training dataset, we are left with twelve configurations to analyze. We label them in the similar fashion as \texttt{A\_0147}, where the first letter (either \texttt{A} or \texttt{B}) stands for the CNN architecture type, and the trailing series of digits corresponds to the input channels used during training. This section first presents the overall CNN performances (Section~\ref{subsec:cnn_perf}), and then focuses on the diagnosis of a chosen model (Section~\ref{subsec:best_model}).

\subsection{Overall CNN performances}
\label{subsec:cnn_perf}

\autoref{fig:roc_curve} shows the receiver operating characteristic curve (ROC curve) for the twelve configurations under study. The area under the curve (AUC) is given in parenthesis next to each configuration name. The AUC is to be interpreted as the probability that given a random observation from class 0 (stable simulation) and a random observation from class 1 (unstable simulation), the CNN outputs a higher value for the later. In other words, it is the probability that the CNN tells which one is which. The right panel of \autoref{fig:roc_curve} is a closer look at the upper left corner of the left panel.

First, we see that both CNN architectures are capable of predicting whether a hierarchical triple system is stable or not, provided the beginning of the evolution of its Kepler elements. This result holds for most input channels: only sets \{4, 7\} and \{4, 8\} provide poor classification results. This means that the raw time series of the inner and outer eccentricities are necessary to accurately predict the stability of hierarchical triple systems evolution.

Interestingly, Architecture B performs roughly as good as Architecture A, even though it has much less trainable parameters (cf \autoref{tab:cnn_architectureA} and \autoref{tab:cnn_architectureB}). This indicates that the coupling between input channels is important for predicting the fate of hierarchical triple systems. This also means that the Keplerian elements should not be treated as independent in the analysis of the system evolution.

Let us now focus on the best models, which ROC curves are better shown on the right panel of \autoref{fig:roc_curve}. The information provided by $i_{\rm mut}$ (channel \texttt{4}) seems to provide relevant information as models \texttt{A\_018} and \texttt{B\_018} perform worse than models \texttt{A\_0147}, \texttt{A\_0148}, \texttt{B\_0147} and \texttt{B\_0148}. The redundancy of channels \texttt{0}, \texttt{1} and \texttt{8} does not seem to worsen the performances of the networks, but are unnecessary.

Finally, the best two models both make use of all Keplerian elements, indicating that they are all meaningful to the CNN. It turns out that the best configurations are determined by the selected input channels, rather than by the CNN architecture type.

\begin{figure*}[ht]
    \centering
    \includegraphics[width=1\textwidth]{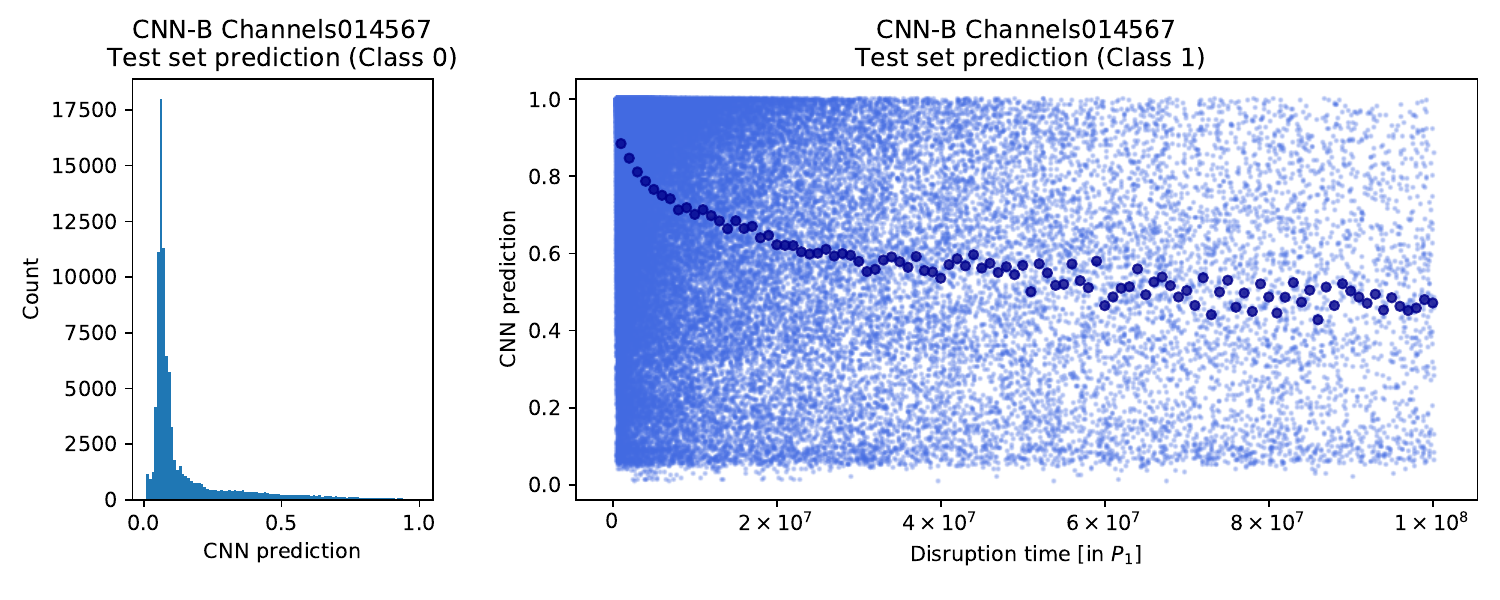}
    \caption{Predictions of the best CNN model \texttt{B\_014567} on the test set. The CNN prediction corresponds to the sigmoid output of the last layer. \textbf{(left)} Distribution of the CNN predictions for the $N=91,456$ stable simulations of the test dataset. The CNN predicted values lie close to zero, meaning it predicts these simulations to be stable. \textbf{(right)} CNN predictions for the $N=88,289$ unstable simulations of the test dataset, as a function of the true breakup time. The CNN predicts values close to one, i.e. unstable simulation. The darker blue dots correspond to the average over 100 equally spaced bins ranging from $5 \times 10^5$ to $10^8 P_1$: the shorter-lived the simulation, the more confident the network.}
    \label{fig:cnnpred_complete_breakup}
\end{figure*}


\subsection{Best model diagnosis}
\label{subsec:best_model}

Although configuration \texttt{A\_014567} has the highest AUC, we prefer to select the configuration \texttt{B\_014567} for parsimony reasons. Given these training input channels, Architecture B has a slightly lower AUC but much less trainable parameters than Architecture A ($59,585$ versus $96,129$). Also, Architecture B allows to take into account stronger dependencies between input channels from the beginning, not permitted by Architecture A.

The left panel of \autoref{fig:cnnpred_complete_breakup} shows the CNN prediction outputs on the $N=91,456$ observations of the test dataset for the stable simulations (class 0). We see that the CNN predicts values close to 0 when it is provided stable simulations. Few stable simulations happen to be more confusing for the network, which outputs values greater than $0.2$. However, studying the stability of hierarchical triple system is not a binary question (stable or unstable?) but rather a quantitative one: how long the system will be stable? We shall interpret the output values between 0 and 1 as the probability that the simulation will not disrupt before $10^8 P_1$.

In the unstable case, the network gives more robust predictions. The right panel of \autoref{fig:cnnpred_complete_breakup} shows a scatter plot of the selected model predictions as a function of the true breakup time for the $N=88,289$ unstable simulations of the test dataset. The CNN mainly predicts values close to 1 for unstable simulations. The scatter plot can be misleading as we can see a lot of points close to the lower end of the x-axis, even on the bottom left corner where we would like to see none. This is because the unstable dataset is biased towards short-lived simulations (see \autoref{fig:disruption_hist_table}), and so is the test subset.

To better understand the prediction performances of the network, we averaged the CNN predictions over 100 equally spaced bins, starting from $5 \times 10^5 P_1$ up to $10^8 P_1$. They correspond to the larger dark blue dots on the right panel of \autoref{fig:cnnpred_complete_breakup}. The average is performed over more observations on the left hand side of the plot than on the right hand side, due to the bias towards short-lived simulations in our unstable dataset. This explains the higher scatter of the averaged predictions at larger disruption time. Nevertheless, we can see that the network predicts values very close to 1 for short-lived systems. But the averaged predictions become smaller as the system lives longer. In other words, the network gives very robust predictions for short-lived systems, and less robust predictions when the system lasts longer.


\section{Discussion}
\label{sec:discussion}

\subsection{Generalization to arbitrary hierarchical triple systems of equal masses}

Our results have been derived from a dataset comprised of hierarchical triple systems with fixed equal masses ($m_1 = m_2 = m_3 = 10 \rm\,M_\odot$). Because of the scale-free nature of Newtonian gravity, any global rescaling of the masses (say, $m_1 = m_2 = m_3 = 30 \rm\,M_\odot$) can be absorbed by rescaling the time unit. Numerical experiments also confirmed that the disruption time of hierarchical triples can be expressed in units of the inner orbital period, without loss of generality, even for systems with different mass or length scales \citep[e.g.][]{hayashi2022}. This consideration is not valid anymore when one includes non-Newtonian forces, such as tides and general relativity corrections, which can affect the evolution of triple systems.

The code and the CNN trained models presented in this works are provided as an open-source package. Currently, the code is optimized to work with \textsc{tsunami} simulation outputs \citep{trani2019b}, but it can be  easily extended to other codes. Because of the above considerations, the body masses have to be the same (but not necessarily $10 \rm\,M_\odot$), as long as the time-series is generated in units of the inner orbital period $P_1$.


\subsection{Limitations}

The CNNs we have presented are set up to receive orbital inputs on a fixed timestep in units of initial inner orbital periods.
Using the initial inner orbital period as time unit may not necessarily be an optimal choice. Indeed, the long-term evolution of the system may lead to small changes in the inner orbital period. As long as the changes even out (c.f. channel 2 on \autoref{fig:data_example} for example) the unit system remains valid. But if these changes happen to become large and consistent, the time unit system will not be meaningful anymore as the mean of $P_1$ will be different. In principle, this could degrade the performances of the artificial neural network.

That said, large changes in the evolution of the inner orbital period also mean that the system becomes very unstable. Since we are interested in the long term stability of the system, we would not need the help of a network in such cases, given that the system is obviously unstable. Our simulations have been selected to lie in the gray area between stability and instability. We therefore expect that the chosen time unit system does not significantly deteriorate the CNN predictions.

\REV{
During the review process of this manuscript, \citet{hamers2022_stability} presented a similar work on estimating the stability of triple systems using machine learning. The main difference from our work is that they estimate the stability of triples only using a limited set of initial conditions (masses, semimajor axes, eccentricities and mutual inclinations), whereas we take into consideration the full set of initial parameters. By obtaining the early evolution of the triples, we can predict the stability for any individual realization of a hierarchical triple system.
On the other hand, any stability criterion bases only on partial initial conditions, like the analytic \citet{mardling2001} criterion or the neural network by \citet{hamers2022_stability}, neglect parameters like true anomalies and the apsidal orientations. $N$-body simulations have shown that these parameters also determine the long-term stability of a triple and can turn stable triples into unstable ones and vice-versa \citep[e.g.][]{hayashi2022}. This is because triple systems are inherently chaotic, and the boundary between stable and unstable orbits is fractal in the initial phase space. Consequently, any criterion based only on partial initial conditions is destined to be approximate at best.

Among the other differences with the work of \citet{hamers2022_stability}, it is worth noting that their criterion is based on the short-term evolution of the triples, because they stop the simulations at $t_{\rm f} = 100 \, P_2$, which is the minimum allowed duration of the simulations we consider. Additionally, we fully explore the parameter space in initial conditions, while they explore only slices of the initial parameter space, which could result in a biased training data.
}

\subsection{Conclusion \& future work}

Hierarchical triple systems are of major importance in astrophysics: from the study of planetary systems evolution to the formation of gravitational waves. These systems are mainly studied through complete N-body integration, which can be computationally expensive. Instead, we propose to use a CNN to predict the stability of hierarchical triple system using only $0.5\%$ of the total integration time (therefore being $200$ times faster).

We have shown that the time-series evolution of the Keplerian elements allows to accurately predict the long term survival of hierarchical triple systems. In particular, the eccentricities provide relevant information for stability prediction. Our selected CNN model has an AUC of 0.956, making it robust for classification on new unseen data.

Our model is computational efficient and can be incorporated into more complex models triple system evolution \citep[e.g.][]{toonen2021,hamers2021}, or it can be used to put stability constraints on observed systems. We provide all trained models, their diagnosis and the code in an open source GitLab repository\footnote{\texttt{https://gitlab.com/aatrani/triple-stability-classifier}}.

\REV{Possible extensions of our work include the generation of a larger training dataset involving very short-lived simulations and non-equal mass systems, larger and deeper convolutional architectures, an optimized training scheme with a train-validation-test data splits. We leave these developments to future works.}


\section*{acknowledgments}
A.~A.~T. acknowledges support from JSPS KAKENHI Grant Numbers 17H06360, 19K03907 and 21K13914. The simulations were run on the Deigo high performance computing cluster at OIST. We are grateful for the help and support provided by the Scientific Computing section at OIST.
F.~L. wishes to thank Kenji Doya and Jun Makino for stimulating discussions.
We finally thank the anonymous referees for their careful reading of our manuscript, and their constructive feedback, suggestions, and questions.


\appendix

\section{Two CNN architectures}
\label{app:cnn_architectures}

We introduce two convolutional neural network architectures to account for different dependencies between input channels and assess their significance regarding binary classification performances.

Architecture A uses convolution filters of shape $(1, 5)$ exclusively. These filters slide along the time axis of each input channel independently. They are shared by all input channels, which means that the CNN starts looking for patterns in each input channel independently. Only the last two layers are fully connected and allow for potential dependencies between input channels. We alternate convolutional layers with max-pooling layers of size 4, and we repeat this pattern six times. The six convolutional layers have respectively 16, 32, 32, 64, 64 and 64 filters, to allow for a reasonable amount of parameters (see \autoref{tab:cnn_architectureA}).

Architecture B, on the other hand, starts with a convolution filter of size $(c, 5)$ across all input channels. This large filter transforms the multiple input time series into one single output time series (a sort of mixture of all input channels). This allows to account for stronger dependencies between input channels from the beginning. The large convolution has 64 output filters, which is flexible to consider several kinds of dependencies from the beginning. The rest of Architecture B is similar to Architecture A. Due to its design, Architecture B has quite less trainable parameters than Architecture A (see \autoref{tab:cnn_architectureB}).

\begin{table}[ht]
\parbox{.48\linewidth}{
\centering
\begin{tabular}{|c|c|c|c|}
\hline
\textbf{Layer}          & \textbf{Output shape} & \textbf{Nb params.}  \\ \hline
\texttt{Input}          & (c, 31830, 1)         & 0                    \\ \hline 
\texttt{Conv2D}         & (c, 31826, 16)        & 96                   \\ \hline 
\texttt{MaxPooling}     & (c, 7956, 16)         & 0                    \\ \hline 
\texttt{Conv2D}         & (c, 7952, 32)         & 2592                 \\ \hline 
\texttt{MaxPooling}     & (c, 1988, 32)         & 0                    \\ \hline
\texttt{Dropout}        & (c, 1988, 32)         & 0                    \\ \hline
\texttt{Conv2D}         & (c, 1984, 32)         & 5152                 \\ \hline 
\texttt{MaxPooling}     & (c, 496, 32)          & 0                    \\ \hline 
\texttt{Conv2D}         & (c, 492, 64)          & 10304                \\ \hline 
\texttt{MaxPooling}     & (c, 123, 64)          & 0                    \\ \hline
\texttt{Dropout}        & (c, 123, 64)          & 0                    \\ \hline
\texttt{Conv2D}         & (c, 119, 64)          & 20544                \\ \hline 
\texttt{MaxPooling}     & (c, 29, 64)           & 0                    \\ \hline 
\texttt{Conv2D}         & (c, 25, 64)           & 20544                \\ \hline 
\texttt{MaxPooling}     & (c, 6, 64)            & 0                    \\ \hline
\texttt{Dropout}        & (c, 6, 64)            & 0                    \\ \hline
\texttt{Flatten}        & 384c                  & 0                    \\ \hline
\texttt{FullyConnected} & 16                    & 6144c + 16           \\ \hline
\texttt{FullyConnected} & 1                     & 17                   \\ \hline
\end{tabular}
\caption{CNN Architecture A. This architecture treats each time series separately until the last three layers, where the channels are flatten and the information is blended. The convolution masks are shared across all channels. The total number of parameters is $59,265 + 6,144c$, where $c$ is the number of channels in the input time series.}
\label{tab:cnn_architectureA}
}
\hfill
\parbox{.48\linewidth}{
\centering
\begin{tabular}{|c|c|c|c|}
\hline
\textbf{Layer}          & \textbf{Output shape} & \textbf{Nb params.}  \\ \hline
\texttt{Input}          & (c, 31830, 1)         & 0                    \\ \hline 
\texttt{Conv2D}         & (1, 31826, 64)        & 320c + 64            \\ \hline 
\texttt{MaxPooling}     & (1, 7956, 64)         & 0                    \\ \hline 
\texttt{Conv2D}         & (1, 7952, 32)         & 10272                \\ \hline 
\texttt{MaxPooling}     & (1, 1988, 32)         & 0                    \\ \hline
\texttt{Dropout}        & (1, 1988, 32)         & 0                    \\ \hline
\texttt{Conv2D}         & (1, 1984, 32)         & 5152                 \\ \hline 
\texttt{MaxPooling}     & (1, 496, 32)          & 0                    \\ \hline 
\texttt{Conv2D}         & (1, 492, 32)          & 5152                 \\ \hline 
\texttt{MaxPooling}     & (1, 123, 32)          & 0                    \\ \hline
\texttt{Dropout}        & (1, 123, 32)          & 0                    \\ \hline
\texttt{Conv2D}         & (1, 119, 64)          & 10304                \\ \hline 
\texttt{MaxPooling}     & (1, 29, 64)           & 0                    \\ \hline 
\texttt{Conv2D}         & (1, 25, 64)           & 20544                \\ \hline 
\texttt{MaxPooling}     & (1, 6, 64)            & 0                    \\ \hline
\texttt{Dropout}        & (1, 6, 64)            & 0                    \\ \hline
\texttt{Flatten}        & 384                   & 0                    \\ \hline
\texttt{FullyConnected} & 16                    & 6160                 \\ \hline
\texttt{FullyConnected} & 1                     & 17                   \\ \hline
\end{tabular}
\caption{CNN Architecture B. This architecture begins with a shared convolutional layer which blends all input channels into a single time series. The rest of the architecture therefore becomes one-dimensional. The total number of parameters is $57,665 + 320c$, where $c$ is the number of channels in the input time series.}
\label{tab:cnn_architectureB}
}
\end{table}

The input layer of both CNN architectures has a length of $31,830$. This value corresponds to $t_f / \Delta t$ and is the same for all simulations.
\REV{Also note that we employ \texttt{Conv2D} filters (and not \texttt{Conv1D}) even though we do not use the first dimension of the filter in practice. Indeed, most filters have size $(1, 5)$, where the first dimension allows to input different channels. The first layer of Architecure B employs a convolutional filter of size $(c, 5)$ which uses the first dimension to blend all time-series together.}

\section{Activation functions}
\label{app:activation_functions}

For both network architectures, all activation functions are Rectified Linear Units (ReLU), except for the activation function between the last two fully connected layers which is a sigmoid function. The ReLU activation function defined as
\begin{equation}
f(x) = \max(0, x)
\end{equation}
is standard in artificial neural networks and is used to break the linearity between subsequent layers. The sigmoid activation function is defined as
\begin{equation}
\sigma(x) = \frac{e^x}{1 + e^x}
\end{equation}
and outputs a value between 0 and 1, which can be interpreted as the probability for the input data of being from class 1 (unstable simulation) rather than class 0 (stable simulation).

\section{Training strategy}
\label{app:training_strategy}

Both CNN architectures are trained with the same strategy. As it is standard for binary classification tasks, we use the binary cross-entropy loss function to assess the network training performances:
$$
\mathcal{L} = \frac{1}{N_b} \sum_{i=1}^{N_b} y_i \log(p_i) + (1 - y_i) \log(1 - p_i)
$$
where $N_b$ is the batch size, $y_i \in \{0, 1\}$ is the true label for observation $i$ in the dataset (1 for unstable simulation, 0 for stable simulation), and  $p_i \in [0, 1]$ is the output of the softmax activation at the last layer of the CNN.

The loss function is optimized with respect to the network parameters. The network parameters are the convolutional filters coefficients. The optimization is perform with the Adam optimizer with default values: \texttt{Adam(learning\_rate=0.001, beta\_1=0.9, beta\_2=0.999, epsilon=1e-07)}.

During training, we use a generator yielding batches of size $N_b = 64$ observations from the training dataset. The training generator successively draws individual observation, with equal probability of coming from the stable or the unstable training dataset. One training step corresponds to one update of the network parameters using the Adam optimizer over the batch of size $N_b$. We perform $50$ training steps per epoch, which amounts to $3,200$ observations seen by the network at every epoch. Recall that the training dataset has $2 \times 150,000 = 300,000$ observations. All CNN models have been trained for 250 epochs.


\bibliography{sample631}{}
\bibliographystyle{aasjournal}



\end{document}